# Security-Performance Tradeoffs of Inheritance based Key Predistribution for Wireless Sensor Networks

Rajgopal Kannan, Lydia Ray, Arjan Durresi and S. S. Iyengar
Department of Computer Science
Louisiana State University
Baton Rouge, LA 70803

## Abstract

Key predistribution is a well-known technique for ensuring secure communication via encryption among sensors deployed in an ad-hoc manner to form a sensor network. In this paper, we propose a novel 2-Phase technique for key predistribution based on a combination of inherited and random key assignments from the given key pool to individual sensor nodes. We also develop an analytical framework for measuring security-performance tradeoffs of different key distribution schemes by providing metrics for measuring sensornet connectivity and resiliency to enemy attacks. In particular, we show analytically that the 2-Phase scheme provides better average connectivity and superior $q$-composite connectivity than the random scheme. We then prove that the invulnerability of a communication link under arbitrary number of node captures by an adversary is higher under the 2-Phase scheme. The probability of a communicating node pair having an exclusive key also scales better with network size under the 2-Phase scheme. We also show analytically that the vulnerability of an arbitrary communication link in the sensornet to single node capture is lower under 2-Phase assuming both network-wide as well as localized capture. Simulation results also show that the number of exclusive keys shared between any two nodes is higher while the number of $q$-composite links compromised when a given number of nodes are captured by the enemy is smaller under the 2-Phase scheme as compared to the random one.

## I. INTRODUCTION

Sensor networks are autonomous systems of tiny sensor nodes equipped with integrated sensing and data processing capabilities. They can be deployed on a large scale in resource-limited and harsh environments such as seismic zones, ecological contamination sites or battlefields [1], [8]. Their ability to acquire spatio-temporally dense data in hazardous and unstructured environments makes them attractive for a wide variety of applications [4], [17], [21]. Sensor networks (sensornets) are distinguished from typical ad-hoc wireless networks by their stringent resource constraints and larger scale. These operational constraints impose severe security challenges since sensornets may be deployed in hostile environments where nodes are subject to capture and communication links are subject to monitoring [13], [18], [23], [22], [16].

Nodes in a sensornet are typically deployed in an ad-hoc manner into arbitrary topologies before self-organizing into a multihop network for collecting data from the environment and forwarding to the base station or sink [1],[5]. Establishing a secure communication infrastructure among a collection of arbitrarily deployed sensor nodes is an important and challenging security issue (known as the *bootstrapping problem* [2]). Due to severe computational and memory constraints, symmetric key cryptography is the most feasible encryption mechanism for node to node communication.

This work was supported by NSF grants, IIS-0312632 and IIS-0329738 and DARPA and AFRL grant # F30602-02-1-0198.

However the high energy-cost of routing makes traditional methods of key exchange and key distribution protocols based on trusted third party mechanisms difficult to implement.

Since bootstrapping should not rely on pre-existing trust associations between fixed sensor nodes or the availability of an on-line service to establish these trust associations, an attractive alternative for secure encrypted communication between adjacent sensor nodes is key predistribution, i.e. pre-installing a limited number of keys in sensor nodes prior to actual deployment. Key predistribution is also challenging since ad-hoc network deployment makes it impossible to pre-determine the neighborhood of any node, yet key distribution schemes must *ensure good network connectivity*(through key sharing) and *resilience to node/key capture by the enemy* even with limited number of keys per node. A trivial predistribution solution is to have a single secret key shared among all nodes. While this solution keeps the network fully connected (every node can communicate with every other node) and scalable (new nodes can be added without any keying overhead), it provides extremely poor resiliency to enemy attack. At the opposite end of the spectrum, one can have each pair of nodes sharing a distinct key. This solution provides both high connectivity and high security but is very memory-intensive and not scalable.

There have been several recent works on key pre-distribution [6], [2], [14], [3], [11]. The pioneering paper in [6] proposes a simple, scalable probabilistic key predistribution scheme in which a certain number of keys are drawn at random from a (large) key pool and distributed to sensor nodes prior to their deployment. Post-deployment, adjacent nodes participate in shared key discovery. A logical graph is created in which edges exist between adjacent sensor nodes sharing at least one key. This is followed by the establishment of paths between nodes using secure links in the logical graph. In [2], the authors have presented new mechanisms for key establishment using the random key pre-distribution scheme of [6] as a basis. Their $q$-composite scheme requires that two adjacent communicating nodes have at least $q$ keys in common. This scheme provides high resiliency against small scale enemy attack.

Note that due to the random distribution of keys and adhoc deployment of sensors, there is a non-negligible probability of a disconnected logical graph. The degree of connectivity of the resultant sensor network under a given key predistribution scheme is therefore an important performance metric. There is also a strong correlation between network connectivity and security. Adversaries that capture nodes can gain complete information about the keys stored at the node in the worst case. Thus in order to make the network less vulnerable to node/key capture the overall key pool size must be large. Since individual sensor nodes have limited memory for key storage, this reduces the probability of having a large number of shared keys between neighboring sensors.

Good solutions for key pre-distribution must be memory-efficient and scalable, simultaneously ensuring that (a majority of) the network is connected through secure communication links **and** provide high resiliency to enemy attack so that the capture of a few sensor nodes does not (severely) compromise network communication. In this paper, we propose a novel solution to the key predistribution problem (labeled 2-Phase key predistribution) that exploits the connectivity and capture-resiliency properties of loading sensor nodes with a combination of *randomly derived* and *inherited* keys We evaluate our solution by analytically developing novel quantitative metrics that measure the key predistribution schemes' security-performance tradeoffs in terms of the network resiliency to node/key capture, the number of available secure links and the key (memory) requirement per node for a given level of connectivity. We compare the network connectivity and security performance and show analytically and through simulations that the proposed 2-Phase scheme strongly favors highly secure large-composite key communication and is more resilient to node capture than the random scheme. We first show analytically that the invulnerability of an arbitrary $q$-composite communication link to any number of node captures is higher in our scheme. We also derive analytical results for measuring the vulnerability of a $q$-composite link to single-node capture assuming adversaries who can use captured-key knowledge network-wide as well as locally and show that the 2-Phase scheme is more resilient. Finally, we present simulation results that show the number of exclusive keys shared between two nodes is higher while the number of $q$-composite links compromised when a given number of nodes are captured by the enemy is smaller under the 2-Phase scheme.

The paper is organized as follows: Section 2 describes some of the related work on key predistribution. Section 3 describes the proposed 2-Phase scheme. Section 4 outlines our metrics for measuring security-performance tradeoffs. Section 5 contains some analytical connectivity results with Section 6 containing analytical results on security of communication edges under node capture. Section 7 describes simulation results followed by some implementation issues and Conclusions in Section 9.

## II. RELATED WORK: OVERVIEW OF THE BASIC RANDOM KEY DISTRIBUTION SCHEME

In general, key management for sensor networks consists of three phases, key pre-distribution, shared key discovery and path establishment. There have been several recent works on key pre-distribution [6], [2], [14], [10], [11]. The pioneering paper in [6] proposes a simple probabilistic key pre-distribution scheme which works as follows: A pool of $L$ keys with key identifiers is generated. For each node, $k << L$ keys are drawn at random from the pool and are installed into the memory of the node. The *shared key discovery* phase takes place after the deployment of sensor nodes. Each node broadcasts its key identifiers. After discovering a shared key with a neighboring node, a node can verify that the neighbor actually holds the key through a challenge-response protocol. A link is then established using the shared key. A (logical) graph with secured communication links is formed after this phase. Due to random distribution of keys and ad-hoc deployment of sensors, there exists a chance that the logical graph is disconnected, in which case sensor nodes perform *range extension* [2] until a connected logical graph is created. However, in terms of node energy, this procedure is quite expensive for a sensornet.

In [2], the authors have presented new mechanisms for key establishment using the random key pre-distribution scheme of [6] as a basis. Their $q$-composite scheme requires that two adjacent communicating nodes must have at least $q$ keys in common. This scheme provides high resiliency against small scale enemy attack.

In [14], the authors evaluate the random key predistribution scheme under a variety of non-random node deployment probability distributions. However, for many realistic sensor networks as visualized in Smart Dust [19], node deployment into known topologies is an unrealistic assumption. [3] presents an alternative model for secure sensor communication using polynomials rather than keys but the computational constraint of using such polynomials is not extensively evaluated.

For sensornets with significant memory constraints, [12] proposes a *deterministic* subvector based key predistribution approach based on distributed agreement using vector spaces and quorums [15]. The scheme has the deterministic property that any two nodes sharing a key under a given mapping (out of several possible mappings) from sensor nodes to keys share exactly two keys (that are unique to these two nodes) in this mapping. For a given network deployment, two physically adjacent sensors can encrypt messages using shared keys under one or more of these mappings, where each mapping yields a 2-composite key. It can be shown that any adhoc deployment of sensor nodes yields very high connectivity with low key storage requirements per node.

## III. Two-Phase Key Predistribution Mechanism

We now describe a novel key pre-distribution scheme (labeled 2-Phase) in which sensor nodes are preloaded with a combination of *randomly derived* and *inherited* keys. Our proposed 2-Phase scheme is motivated by the following key observations:

- From the connectivity point of view, the probability of having a common key between two nodes decreases as the key pool size increases under the random pre-distribution scheme. We observe that the (probabilistic) connectivity of the logical graph can be increased if we can ensure that each node *deterministically* shares some of its keys with some nodes (as in the subvector scheme [12]).

- We hypothesize that it is better from the security point of view to pre-distribute keys in a less-random fashion such that whenever a node shares a key with another node, it should be likely to share a larger number of keys with this node, If so, the resulting network should consist of high-composite links. Note that $q$-composite schemes are more secure with increasing $q$. If the adversary has obtained $X$ keys (through the capture of one or more sensor nodes), the probability of determining the exact $q$-subset of $X$ that is used by a given communicating sensor pair decreases exponentially with increasing $q$.

We now describe the key steps in the proposed 2-Phase key predistribution mechanism. Order the sensor nodes apriori in a logical queue and distribute keys in increasing order according to the rules below.

- The first node is assigned $k$ keys drawn randomly from the key pool of size $L$.

- For every succeeding sensor node $i$, $k$ keys are distributed in two consecutive phases. First, node $i$ receives a predetermined fraction $f$ $(1/k \le f < 1)$ of its $k$ keys drawn randomly from the key space of node $i-1$. The remaining $(1-f)$ fraction of $k$ keys are then drawn randomly from the key pool of size $L-k$, *after excluding all $k$ keys of node $i-1$ from $L$.*

The 2-Phase scheme is designed to be biased in favor of nodes sharing several keys with their immediate predecessors and successors, through direct inheritance as well as a random component. Intuitively, this key predistribution methodolody should offer better secure connectivity in the logical graph by inducing the sharing of larger number of keys between nodes, thereby enabling $q$-composite communication for larger values of $q$. More suprisingly however, as we show in the security analysis section, this methodolgy also provides enhanced security under node capture/eavesdropping by allowing for more 'exclusive' key sharing between communicating nodes. The fraction $f$ (called *inheritance ratio*) plays a significant part in the connectivity/security of the logical graph created after node deployment. Note that the random key predistribution scheme **is not** a special case of the 2-Phase scheme with $f = 0$, since we eliminate all $k$ keys of the previous node from regardless of the value of $f$. We will shortly derive relationships between 'good' values of the various parameters $k$, $L$, $f$ etc. Finally, the proposed 2-Phase scheme is scalable since new sensor nodes can be assigned keys according to this rule at any time.

Note that there is an implicit ordering of sensors based on their position in the logical queue which determines each nodes key set. Thus each node has a logical identifier which we will refer to as its LID. Storing a node's LID in memory is an implementation decision as there is an associated security-performance tradeoff. If LID's are stored, nodes can be restricted to forming communication links only with adjacent nodes whose LIDs are greater than a specifed minimum and within a specified maximum LID distance. As shown later, this will encourage the formation of high-composite encrypted communication that are also less vulnerable to compromization in the case of node capture. Conversely, storing LIDs will enable the adversary to target nodes with specific LIDs (although their positions will still be unknown). Therefore this becomes an implementation issue.

## IV. Metrics for Measuring Security-Performance Tradeoffs

Since security mechanisms directly impact system performance, there is a strong need to develop a rigorous analytical framework for measuring the security-performance tradeoffs of arbitrary key distribution schemes. These tradeoffs can be represented as functions of individual metrics which measure the networks 'secure' connectivity in terms of the number of available secure links or paths, the memory requirement in terms of keys per node for a given level of connectivity and measuring resiliency of the network to node/key capture. In this paper, we obtain some new analytical results on the security-performance tradeoffs of key predistribution schemes using the quantitative metrics outlined below. Results for the proposed 2-Phase scheme are compared with random key predistribution.

- Connectivity Metrics

  - *Logical sensor degree*: We measure the logical degree of a node as the number of adjacent sensor nodes (in the logical graph) with which it shares at least one key. The higher the expected node degree, the better the connectivity of the logical graph. A high expected degree also implies a larger expected number of disjoint paths from any source to any destination. Multiple disjoint paths can be used to split communication and carry disjoint messages, thereby increasing overall data security. We show that nodes under the proposed 2-Phase scheme have higher expected degrees as compared to random key predistribution.

  - *Number of keys shared between any two neighboring nodes*: This metric can be used to evaluate connectivity under $q$-composite key communication. We show that any two sensor nodes are expected to share more keys and are more likey to share $q$ keys for any value of $q$ (thereby enabling $q$-composite communication), as compared to random key predistribution.

- Security Metrics

  - *Exclusive Key Sharing*: If two communicating nodes share one or more keys exclusively, then their communication is *invulnerable* to any number of node captures. Note that the exclusivity metric can be computed network-wide or with respect to a local cluster[1]. Network wide exclusivity between communicating nodes implies resilience against a powerful adversary who can capture nodes and use the captured key information *anywhere* in the sensor network. Alternatively, we can consider a weaker adversary who can use the key information only within the cluster of the captured node.

  - *Node Capture*: We measure the impact of node capture on network security by considering the number of communication links that are no longer secure (i.e only use keys from the captured key pool). We analytically determine bounds on the inheritance ratio $f$ for which the 2-Phase scheme shows good resilience to network-wide as well as localized single-node capture and present simulation results that show good network resilience to multiple-node capture as well. The expected number of links compromised in these cases is shown to be lower for the 2-Phase scheme as compared to the random scheme.

## V. Secure Network Connectivity: Analytical Results

*Proposition 1:* Let $l$ and $i > l$ be any two nodes in the sensornet. The expected number of keys shared by $l$ and $i$ under the 2-Phase and Random schemes, respectively, are

$$
\begin{aligned}
E_{l,i}^{2P} &= k\left(\frac{k}{L} + (\frac{fL-k}{L-k})^{i-l}(1-\frac{k}{L})\right) \\
E_{l,i}^{Rand} &= \frac{k^2}{L}
\end{aligned}
$$

*Proof:* The number of common keys between any two nodes under the random scheme is the standard hypergeometric distribution with parameters $k$ and $L$, whose mean is $k^2/L$. For the 2-Phase scheme, let $X_r$ be the number of keys in common between nodes $l$ and $l+r$. Then we have,

$$
\begin{aligned}
X_{r+1} &= fX_r + (k - X_r)\frac{k - fk}{L - k} \\
&= X_r\frac{fL - k}{L - k} + k\frac{k - fk}{L - k}
\end{aligned}
$$

since after selecting an expected $fX_r$ keys from the previous node, there are $k - X_r$ keys of node $l$ left in the random keypool of the current node. $E_{l,i}^{2P} = X_{i-l}$ is the solution to the above recurrence relation with intial condition $X_0 = k$. $E_{l,i}^{2P} > E_{l,i}^{Rand}$ as expected. ∎

---

[1]Typical sensor networks are organized into hierarchical clusters with cluster heads, such that each node is within wireless range of other nodes in the cluster [7]. Thus a compromised node can potentially eavesdrop on all intra-cluster communication.

Thus to ensure $q$-composite connectivity between arbitrary nodes, a good choice is to select $k$ and $L$ such that $q = k^2/L$. Further, if $f = k/L$, then the expected number of common keys between any two nodes is identical under both schemes.

*Corollary 1:* The probability that any two nodes share at least $q$ keys and the expected $q$-composite degree of a sensor node (i.e., number of neighbors with which it shares more than $q$ keys) is higher under the 2-Phase key distribution scheme, $f \geq k/L$.

As nodes are more likely to share multiple keys under 2-Phase, the probability of uncovering all such common keys (which is necessary to decipher data transmissions between the two nodes) can be shown to be lower and hence two-phase is more secure in this respect.

## VI. Network Resiliency against Enemy Attack: Analytical Results

In this section, we propose some quantitative metrics for measuring the security of communication links under enemy attack and analytically evaluate these metrics under different adversarial models. We assume an adversary that is able to capture nodes and obtain full knowledge of the captured node's key space. We evaluate link security under a 'network-wide' adversary who can use knowledge of captured keys to compromise communication in any part of the network (regardless of the physical location of the captured node). Our results can be easily extended to analyze link vulnerability in the presence of a localized adversary who utilizes captured key knowledge locally, i.e. can compromise communication within a small neighborhood of the captured node (for example, its cluster as in LEACH [7]).

### A. Vulnerability Under Multiple Node Capture: Key Exclusivity

We first evaluate the vulnerability of logical communication links in the sensornet to multiple node capture. An obvious metric for measuring this vulnerability is the degree of exclusivity of the keys used by any two neighboring nodes for setting up a communication link. Therefore we evaluate the probability of any two neighboring nodes containing exactly one *network-wide* exclusive key, the presence of which will render their communication link invulnerable to any number of (other) node captures [2].

*Proposition 2:* **Key Exclusivity**: In an $N$ node sensor network, the probability that a given communication link between two arbitrary neighboring sensors is invulnerable to any number of network-wide node captures is given by:

$$\text{2-Phase:} \quad \frac{(1-f)^4}{1-(k/L)} \left(\frac{k}{L}\right)^2 \left(1 - \frac{k}{L}\left(\frac{1-f}{1-(k/L)}\right)\right)^{N-5}$$

$$\text{Random:} \quad \left(\frac{k}{L}\right)^2 \left(1 - \frac{k}{L}\right)^{N-2}$$

---

[2]In general, we can compute the probability of two nodes containing at least one exclusive key, but for all practical purposes this probability drops off extremely rapidly for more than one exclusive key. Hence we obtain a simple lower bound on invulnerability by focusing on the presence of a single exclusive key.

Link invulnerability is higher under the 2-Phase scheme scheme for $\frac{1}{k} \leq f \leq \frac{k}{L}$.

*Proof:* Let $IV^{rand}$ and $IV^{2P}$ denote the probability that two arbitrary neighboring sensor nodes $i$ and $j$ communicate using an exclusive key under the two key predistribution schemes. In the case of the 2-Phase scheme, $i$ and $j$ represent the LIDs of the communicating nodes. Consider a specific key $a$ from the key pool.

For the random scheme, the probability that both nodes $i$ and $j$ possess key $a$ is $(k/L)^2$ while the probability that an arbitrary node $l \neq \{i, j\}$ does not possess key $a$ is $1 - \left( \binom{L-1}{k-1} / \binom{L}{k} \right) = 1 - k/L$. Hence the invulnerability of the link between nodes $i$ and $j$ under any number of node captures is given by:

$$IV^{Rand} = \left( \frac{k}{L} \right)^2 \left( 1 - \frac{k}{L} \right)^{N-2} \tag{1}$$

For the 2-Phase scheme, the probability that key $a$ is exclusive to nodes $i$ and $j$ is the probability that node 1 does not select key $a$, followed by all nodes up to node $i - 1$ not selecting key $a$ conditioned on the fact that their predecessor node did not select key $a$. Node $i$ then selects key $a$ given that node $i - 1$ did not select it. Similarly all nodes after $i$ conditionally do not select key $a$ except node $j$.

Let $P(1^c)$ denote the probability that node 1 does not contain key $a$, $P(1^c) = 1 - k/L$ since node 1 selects keys from the keypool first. Similarly, let $P(l^c \mid (l-1))$ denote the probability that a node $l$ does not contain key $a$ given node $l - 1$ contains it, $P(l^c \mid (l-1)) = 1 - f$, by definition. Finally, we have

$$
\begin{aligned}
P(l^c \mid (l-1)^c) &= \frac{\binom{L-k-1}{k-fk}}{\binom{L-k}{k-fk}} \\
&= 1 - \frac{k}{L} \left( \frac{1-f}{1-(k/L)} \right) \\
P(l \mid (l-1)^c) &= \frac{k}{L} \left( \frac{1-f}{1-(k/L)} \right)
\end{aligned}
$$

We now consider two cases (WLOG assume $j > i$):

**Case 1:** $j > i + 1$:

$$
\begin{aligned}
IV^{2P} &= P(1^c) P(2^c \mid 1^c) \cdots P((i-1)^c \mid (i-2)^c) P(i \mid (i-1)^c) P((i+1)^c \mid i) \\
&\quad P((i+1)^c \mid i) \cdots P(j \mid (j-1)^c) P((j+1)^c \mid j) \cdots P(N^c \mid (N-1)^c) \\
&= \frac{(1-f)^4}{1 - \frac{k}{L}} \left( \frac{k}{L} \right)^2 \left( 1 - \frac{k}{L} \left( \frac{1-f}{1-(k/L)} \right) \right)^{N-5} \tag{2}
\end{aligned}
$$

**Case 2:** $j = i + 1$:

$$IV^{2P} = (1-f)^2 f \frac{k}{L} \left( 1 - \frac{k}{L} \left( \frac{1-f}{1-(k/L)} \right) \right)^{N-4} \tag{3}$$

Comparing Equations 2 and 1, the probability of two nodes having a network wide exclusive key (i.e. link invulnerability) is higher under the two-phase scheme as compared to the random scheme for $\frac{1}{k} \leq f \leq \frac{k}{L}$. ∎

We can consider an alternative version of the 2-Phase scheme that provides much greater key exclusivity. The first step of key selection is the same as before, i.e. node $i$ selects $fk$ keys from the key space of node $i-1$. However, in the second step, only the $fk$ keys selected from node $i-1$ are excluded from keypol $L$ before node $i$ selects its remaining $k - fk$ keys. For this modified 2-Phase scheme called 2PWR (2-Phase with replacement), we can show the following:

*Proposition 3:* **Scalable Comparitive Exclusivity**: The invulnerability of a communication edge under any number of node captures when keys are distributed using the 2PWR scheme is

$$IV^{2PWR} = IV^{Rand} \frac{(1-f)^4}{\left(1 - f\frac{k}{L}\right)^{N-1}}$$

Thus link invulnerability under 2PWR outperforms the random scheme as the size of the sensornet $N$ scales upward. This link invulnerability is maximized when

$$f = \frac{(N-1)\frac{k}{L} - 4}{(N-5)\frac{k}{L}}$$

*Proof:* Using the same technique as in proposition 2 the probability of a given communication link $(i, j)$ containing a network-wide exclusive key under the 2PWR scheme is given by:

$$
\begin{aligned}
IV^{2PWR} &= \frac{(1-f)^4}{\left(1 - f\frac{k}{L}\right)^{N-1}} \left(\frac{k}{L}\right)^2 \left(1 - f\frac{k}{L}\right)^{N-2} \\
&= \frac{(1-f)^4}{\left(1 - f\frac{k}{L}\right)^{N-1}} IV^{Rand}
\end{aligned}
\tag{4}
$$

The value of $f$ that maximizes the above term can then be found using elementary calculus. $\blacksquare$

While key exclusivity and (average network connectivity) are superior under the 2PWR scheme, the vulnerability of a link to single node capture is lower under the standard 2-Phase scheme as shown in the next section and hence we focus on that scheme for the rest of the paper. The choice of particular key distribution scheme with its associated security performance tradeoffs then becomes an implementation issue.

The next section contains some analytical results on edge vulnerability under localized node capture when the average node density of the sensornet (i.e size of a network cluster) is $M$. Simulation results on the number of exclusive keys per communicating node pair in a cluster are presented in Section 7.

## B. Link Vulnerability Under Single Node Capture

We now consider the vulnerability of communication links in the sensornet to the capture of a single node by the adversary. We assume that the adversary does not posess any extra knowledge about the network topology and thus the capture of any given node by the adversary is equally likely.

Let $i$ and $j$ be any two communicating sensors in radio range and suppose the adversary captures node $l$. The vulnerability of edge $(i,j)$ is the expectation (over all network nodes) that node $l$ contains all the keys in common between $i$ and $j$. We can thus define a network-wide vulnerability metric $VC$ for arbitrary edges in the sensornet as follows:

$$VC = \sum_{l \neq i,j} P[\text{node } l \text{ is captured}] \cdot P[l \text{ contains all keys used to communicate over } (i,j)] \qquad (5)$$

Before describing our vulnerability results, we first prove the following useful lemmas.

*Lemma 1:* Let $i, i-x, i+x$ be arbitrary nodes in a sensornet in which keys have been predistributed according to the 2-Phase scheme. Let $Z$ be any subset of keys from the keyset of node $i$, $|Z| \leq k$. The probability distribution of the number of keys from $Z$ that appear in nodes $i-x$ and $i+x$ are identical and dependent only on the LID difference $x$ for both 2PWR as well as 2POR.

*Proof:* Clearly, the total number of common keys between nodes $i-1$ and $i$ and between nodes $i$ and $i+1$ follows the same probability distribution, since they are obtained in an identical manner through inheritance followed by keys from the random pool. Thus the number of common keys from any subset $Z$ of $i$'s keys also follows the same distribution in $i-1$ and $i+1$. The lemma follows by induction on $x$. ∎

The following statements follow directly from lemma 1 since the number of keys in common between any two nodes under 2-Phase depends only on their LID difference.

*Corollary 2:* Let $i$ and $j$ be two arbitrary nodes in the sensornet in which keys are predistributed according to the 2-Phase scheme, $j > i$. Consider nodes $i-t$ and $j+t$, $t \geq 1$.

- The number of keys in common between nodes $i-t, i$ and $j, j+t$ follows identical probability distributions.
- Suppose nodes $i-t, i$ ($j, j+t$, resp.), share exactly $\beta$ keys, $0 \leq \beta \leq k$. Then the number of keys from the remaining keyset of $i$ ($j$, resp.) present in node $j$ ($i$, resp.) follow identical probability distributions.

The above statements also hold for nodes $i, i+y$ and $j-y, j$, where $y \leq \lceil (j-i)/2 \rceil$.

*Lemma 2:* Let $l, i$ and $j > i$ be any three nodes in a sensornet, such that $i$ and $l$ share exactly $\beta$ keys, $0 \leq \beta < k$ and $0 \leq l \leq i + \lceil (j-i)/2 \rceil$. Let $Z$ denote the set of remaining keys in node $i$, $|Z| = k - \beta$. $E_Z$, the expected number of keys from $Z$ that are present in $j$ is given by

$$E_Z = \begin{cases} (k-\beta)\left(\frac{k}{L} + (\frac{fL-k}{L-k})^{j-i}(1-\frac{k}{L})\right) & \text{if } l < i \\ (k-\beta)\frac{k}{L}(1-(\frac{fL-k}{L-k})^{j-l}) & \text{if } i < l \leq i + \lceil (j-i)/2 \rceil \end{cases}$$

*Proof:* Let $X_r$ represent the expected number of keys from keyset $Z$ in node $i+r$ (if $l < i$) and $l+r$ (if $i < l < j$). $E_Z$ is obtained by solving the recurrence relation

$$\begin{aligned} X_r &= fX_{r-1} + (k - \beta - X_{r-1})\frac{k-fk}{L-k} \\ &= X_{r-1}(\frac{fL-k}{L-k}) + (k-\beta)\frac{k-fk}{L-k}. \end{aligned}$$

with initial condition $X_0 = k - \beta$, if $l < i$ and $X_0 = 0$, if $i < l \leq i + \lceil (j-i)/2 \rceil$. ∎

Assume that node $l$ is captured and let $CR_l$ be the Bernoulli random variable indicating whether capture of $l$ reveals all common keys between communicating nodes $i$ and $j$. Denote $PCR_l : Pr.[CR_l = 1]$. We now state our first proposition on sensornet vulnerability under single node capture.

*Proposition 4:* The probability of a given communication link between neighboring sensors $i$ and $j$ being compromized by the capture of an arbitrary node $l \neq i, j$ is given by

$$
\begin{aligned}
PCR^{2P}_{i-t} = PCR^{2P}_{j+t} &\leq P^{2P}_{i-t,i,k} + \sum_{\beta=0}^{k} P^{2P}_{i-t,i,\beta} \frac{\binom{L-2k+\beta}{k-fk}}{\binom{L-k}{k-fk}} \left( 1 - f(\frac{k}{L} + B^{j-i}(1 - \frac{k}{L})) \right)^{k-\beta} \\
&\qquad\qquad\qquad\qquad\qquad\qquad \text{for } t \geq 1 \\
PCR^{2P}_{i+t} = PCR^{2P}_{j-t} &\leq P^{2P}_{i,i+t,k} + \sum_{\beta=0}^{k} P^{2P}_{i,i+t,\beta} \frac{\binom{L-2k+\beta}{k-fk}}{\binom{L-k}{k-fk}} \left( 1 - f(\frac{k}{L}(1 - B^{j-i-t})) \right)^{k-\beta} \\
&\qquad\qquad\qquad\qquad\qquad\qquad \text{for } 1 \leq t < \lceil \frac{j-i}{2} \rceil \\
PCR^{Rand}_l &= P^{Rand}_{l,i,k} + \sum_{\beta=0}^{k} P^{Rand}_{l,i,\beta} \frac{\binom{L-k+\beta}{k}}{\binom{L}{k}} \qquad \forall l \neq \{i, j\}.
\end{aligned}
$$

where $P^{2P}_{l,i,\beta}$ and $P^{Rand}_{l,i,\beta}$ denote the probability that nodes $l$ and $i$ share exactly $\beta$ keys under the specified key distribution scheme and $B = \frac{fL-k}{L-k}$.

*Proof:* Without loss of generality, assume $1 \leq l \leq i + \lceil \frac{j-i}{2} \rceil$ and let nodes $l$ and $i$ share exactly $\beta$ keys. Let $Z$ denote the set of remaining keys in node $i$, $|Z| = k - \beta$. Under the 2-Phase scheme, node $j$ can first obtain keys from keyset $Z$ through inheritance from its predecessor, node $j - 1$, and then from the random keypool of size $L - k$ (obtained after removing the $k$ keys of node $j - 1$) Let $jmZ$ be a random variable denoting the number of keys from $Z$ contained in node $j - 1$ and let $P^r_{jmZ} = Pr.[jmZ = r]$. Let $PNCR_l = 1 - PCR_l$ denote the probability that $j$ contains at least one key from keyset $Z$, for diferent values of $\beta$. Therefore, we have

$$
PNCR_l = \sum_{\beta=0}^{k-1} P^{2P}_{l,i,\beta} \sum_{r=0}^{k-r} P^r_{jmZ} \left( \left\{ 1 - \frac{\binom{k-r}{fk}}{\binom{k}{fk}} \right\} + \frac{\binom{k-r}{fk}}{\binom{k}{fk}} \left\{ 1 - \frac{\binom{L-2k+\beta}{k-fk}}{\binom{L-k}{k-fk}} \right\} \right), \tag{6}
$$

where the first term after the inner summation is the probability that at least one out of $r$ keys is inherited by node $j$ while the second term represents the complementary situation in which at least one key from keyset $Z$ is obtained from the random keypool.

Next, using the fact that $\frac{\binom{k-r}{fk}}{\binom{k}{fk}} \geq (1 - f)^r$ and substituting in Equation 6, we have,

$$\begin{aligned}
PNCR_l^{2P} &\geq \sum_{\beta=0}^{k-1} P_{l,i,\beta}^{2P} \sum_{r=0}^{k-\beta} P_{jmZ}^r \left(1 - (1-f)^r \frac{\binom{L-2k+\beta}{k-fk}}{\binom{L-k}{k-fk}}\right) \\
&= \sum_{\beta=0}^{k-1} P_{l,i,\beta}^{2P} \left(1 - \left(\frac{\binom{L-2k+\beta}{k-fk}}{\binom{L-k}{k-fk}} \sum_{r=0}^{k-\beta} (1-f)^r P_{jmZ}^r\right)\right) \\
&= 1 - P_{l,i,k}^{2P} - \left(\sum_{\beta=0}^{k} P_{l,i,\beta}^{2P} \frac{\binom{L-2k+\beta}{k-fk}}{\binom{L-k}{k-fk}} \sum_{r=0}^{k-\beta} (1-f)^r P_{jmZ}^r\right).
\end{aligned} \qquad (7)$$

Therefore we have

$$PCR_l^{2P} \leq P_{l,i,k}^{2P} + \sum_{\beta=0}^{k} P_{l,i,\beta}^{2P} \frac{\binom{L-2k+\beta}{k-fk}}{\binom{L-k}{k-fk}} \sum_{r=0}^{k-\beta} (1-f)^r P_{jmZ}^r , \qquad l \leq \lceil \frac{j-i}{2} \rceil. \qquad (8)$$

We obtain an efficient approximation for $PCR_l^{2P}$ as follows: Let $j-i=y$, $j-l=m$. Let $p = k/L + B^y(1-(k/L))$ if $l < i$ and $p = (1-B^m)k/L$ if $i < l \leq i + \lceil \frac{y}{2} \rceil$, where $B = (fL-k)/(L-k)$. From lemma 2, the expected number of keys from $Z$ present in node $j-1$ is $p(k-\beta)$. For reasonably small values of $k/L$, we can therefore approximate the distribution of random variable $jmZ$ by the standard Binomial distribution $B(k-\beta, p)$ with the same mean $p(k-\beta)$. Therefore, we get

$$\begin{aligned}
PCR_l^{2P} &\leq P_{l,i,k}^{2P} + \sum_{\beta=0}^{k} P_{l,i,\beta}^{2P} \frac{\binom{L-2k+\beta}{k-fk}}{\binom{L-k}{k-fk}} \left(\sum_{r=0}^{k-\beta} (1-f)^r \binom{k-\beta}{r} p^r (1-p)^{k-\beta-r}\right) \\
&= P_{l,i,k}^{2P} + \sum_{\beta=0}^{k} P_{l,i,\beta}^{2P} (\frac{\binom{L-2k+\beta}{k-fk}}{\binom{L-k}{k-fk}})(1-fp)^{k-\beta} \qquad l \leq \lceil \frac{j-i}{2} \rceil.
\end{aligned} \qquad (9)$$

where $p$ is defined as above.

By corollary 2, $PCR_{i-t}^{2P} = PCR_{j+t}^{2P}$, $t \geq 1$ and $PCR_{i+t}^{2P} = PCR_{j-t}^{2P}$ for $1 \leq t \leq \lceil (j-i)/2 \rceil$. This defines $PCR_l^{2P}$ for all values of $l$ as specified in the statement of the proposition [3].

Finally, using the fact that the probability of node $j$ not containing any key from $Z$ is $\binom{L-k+\beta}{k}/\binom{L}{k}$ under random key predistribution, we derive $PCR_l^{Rand}$ as:

$$PCR_l^{Rand} = P_{l,i,k}^{Rand} + \sum_{\beta=0}^{k} P_{l,i,\beta}^{Rand} \frac{\binom{L-k+\beta}{k}}{\binom{L}{k}} \qquad (10)$$

■

We now consider two separate but related issues: First we determine values of $f$ which minimize the probability of a given communication link $(i, j)$ being compromised under the 2-Phase scheme. Given the security-performance tradeoffs, the user may desire a higher level of connectivity than provided by this optimal $f$. Therefore as an alternate performance metric, we determine values of $f$ for which this probability is lower under 2-Phase key predistribution as opposed to Random key predistribution.

---

[3]Henceforth, we will only use $l \leq \lceil (j-i)/2 \rceil$ for the remaining propositions, using this symmetry.

*Proposition 5:* $PCR_l$, the probability of a given link $(i, j)$ in the sensornet being compromised by capture of any node $l$ is minimized by choosing an inheritance factor $f$ that maximizes the expression

$$x(1 - B^t)(1 - x(2f - f^2)) + f(1 - B^t)B^y(1 + fx - 2x)(1 - x)$$

$$\text{for } l = i - t \text{ and } l = j + t \qquad t \geq 1$$

$$x(1 - B^t)\left[1 - x(2f - f^2)\right) - fB^{y-t}(1 + fx - 2x)\right]$$

$$\text{for } l = i + t \text{ and } l = j - t \qquad 1 \leq t \leq \lceil \tfrac{y}{2} \rceil$$

where $x = k/L$, $j = i + y$ and $B = \frac{fL - k}{L - k}$. If communicating nodes $i$ and $j$ are separated by a minimum distance (i.e. $y > c$, where $c$ is a small constant), then $f = \frac{1}{k}$ minimizes this link vulnerability.

*Proof:* We first obtain a simple approximation for Equation 9. From proposition 1, the expected number of keys in common between any two nodes $i$ and $l$ is given by $p_{il} = k(x + B^{|i-l|}(1 - x))$. Therefore for $k << L$, we can approximate the distribution of the number of common keys $\beta$ between $i$ and $l$ by the Binomial $B(k, p_{il}, \beta)$. Now using $\frac{\binom{L - 2k + \beta}{k - fk}}{\binom{L - k}{k - fk}} \approx (\frac{1 - 2x + fx}{1 - x})^{k - \beta}$, we can rewrite Equation 9 for $l \leq \lceil (j - i)/2 \rceil$ as

$$PCR_l^{2P} \quad \leq \quad P_{l,i,k}^{2P} + \begin{cases} \sum_{\beta=0}^{k} \binom{k}{\beta}(p_{il})^\beta \left((1 - p_{il})(\frac{1 - 2x + fx}{1 - x})(1 - fx - f(1 - x)B^y)\right)^{k-\beta} & \text{if } l < i \\ \sum_{\beta=0}^{k} \binom{k}{\beta}(p_{il})^\beta \left((1 - p_{il})(\frac{1 - 2x + fx}{1 - x})(1 - fx + fxB^{j-l})\right)^{k-\beta} & \text{if } i < l < i + \lceil \frac{j-i}{2} \rceil \end{cases}$$

Substituting for $p_{il}$ and further simplifying, we get

$$PCR_l^{2P} \quad \leq \quad P_{l,i,k}^{2P} + \begin{cases} \left(1 - \left[x(1 - B^t)\left(1 - x(2f - f^2)\right) + f(1 - B^t)B^y(1 + fx - 2x)(1 - x)\right]\right)^k \\ \qquad \text{for } l = i - t \text{ and } l = j + t \qquad t \geq 1 \\ \left(1 - x(1 - B^t)\left[1 - x(2f - f^2) - fB^{y-t}(1 + fx - 2x)\right]\right)^k \\ \qquad \text{for } l = i + t \text{ and } l = j - t \qquad 1 \leq t \leq \lceil \frac{y}{2} \rceil \end{cases} \tag{11}$$

$PCR_l^{2P}$ is minimized by maximizing the inner term as stated in the proposition. When $j - i \geq 4$ and $k << L$, this minimum value is obtained at $f = 1/k$. ∎

*Proposition 6:* The probability of a given link being compromised by the capture of an arbitrary node $l \neq i, j$ in the sensornet is lower under the 2-Phase scheme as compared to Random key predistribution for

$$\frac{1}{k} \leq f \leq x\frac{2 - x}{1 + 2x}$$

where $x = k/L$.

*Proof:* We can express $PCR_l^{Rand}$ in Equation 10 as

$$PCR_l^{Rand} = P_{l,i,k}^{Rand} + \left(1 - \frac{k}{L} + (\frac{k}{L})^2\right)^k \tag{12}$$

by using the approximation $\frac{\binom{L-k+\beta}{k}}{\binom{L}{k}} \approx (1 - \frac{k}{L})^{k-\beta}$ and also approximating the hypergeometric distribution of the number of common keys between nodes $l$ and $i$ by the Binomial $B(k, \frac{k}{L})$ (assuming $k << L$).

Now comparing Equations 12 and 11 using $|l - i| = t$ and assuming $j - i \geq 4$, we have

$$PCR_l^{2P} \leq PCR_l^{Rand} \quad \text{iff} \quad x(1 - B^t)\Big[1 - x(2f - f^2)\Big] \geq x(1 - x)$$
$$\implies \Big[1 - (1 - B^t)(2f - f^2)\Big]x > B^t \tag{13}$$

From the above expression, it can be seen that for reasonably large values of $t$ (i.e. the captured node's LID is not too close to $i$ or $j$, $PCR^{2P} < PCR^{Rand}$ for larger values of $f < 1$. However, the worst-case for link compromization $PCR^{2P}$ occurs when $t = 1$ i.e when nodes $l = i + 1$ or $l = j - 1$ are captured. Therefore substituting $t = 1$ above and simplifying, we get $f < x(2 - x - 2f + 3f^2 - f^3)$ which implies $\frac{1}{k} \leq f \leq x\frac{2-x}{1+2x}$. Hence the 2-Phase scheme has lower vulnerability than the Random scheme for $f$ upto $2k/L$. ∎

From Equation 13, we can see that if the captured node is not too close to the communicating nodes (in terms of LID), then the 2-Phase scheme outperforms Random key predistribution for larger values of $f$, which in turn ensures higher connectivity. In particular, if the adversary is restricted to using knowledge of a captured node's keys within a small neighborhood such as its cluster, then we can further minimize link vulnerability to single-node capture by considering a modified 2-Phase scheme in which two neighboring nodes $i$ and $j$ with $\geq q$ keys in common communicate only if there does not exist any other node $l$ in the cluster such that $i < l < j$. Consider a sensor network with average node denisty $M$. Then the expected LID difference between node $i$ and the nearest node (other than node $j$) is $N/M$. We can therefore approximate link invulnerability to single node capture as follows:

*Proposition 7:* In a sensor network with average node density $M$, the probability that a given communication link $(i, j)$ is invulnerable to single-node capture within its cluster is $1 - (1 - PCR_{i-\frac{N}{M}}^{2P})^M$.

Simulation results in the next section illustrate link vulnerability within a cluster for different sensornet parameters.

Finally, for an average adversary with no specific knowledge about the network topology, the probability of capturing any node $l \neq i, j$ is $1/(N - 2)$. The following proposition then directly follows proposition 6.

*Proposition 8:* The vulnerability metric $VC$ of a given communication link in an $N$-node sensor network with parameters $k$ and $L$, is lower if keys are predistributed using the 2-Phase scheme as compared to random key predistribution, $\frac{1}{k} \leq f < x\frac{2-x}{1+2x}$.

## VII. Simulation Results

In this section, we describe some security and performance results based on simulations carried out on a 1000 node sensor network using a key pool $L$ ranging from 8000-10000 keys. The per-node key space $k$ varies from 40-150 keys. We have evaluated the 2-Phase key distribution scheme for $f = 0.5$. Nodes in the simulations are deployed in clusters as in LEACH [7] where the average node density (in a cluster) varies between 20 to 50 nodes. Figure 1 describe some $q$-composite network connectivity metrics while Figures 2–4 describe several sensornet security metrics.

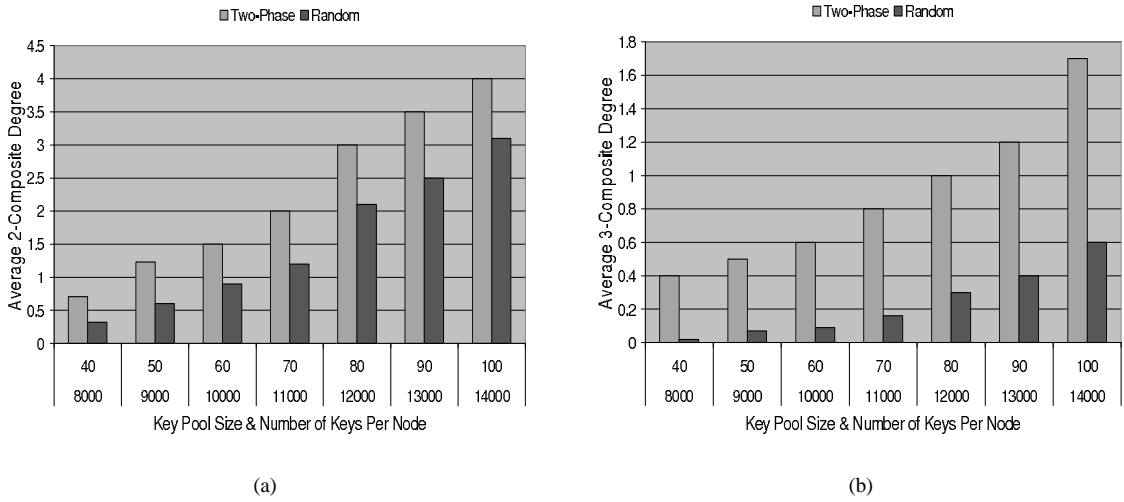

Fig. 1

AVERAGE $q$-COMPOSITE DEGREE OF A NODE (A) 2-COMPOSITE (B) 3-COMPOSITE .

Figure 1 describes the average $q$-composite degree of a node for different values of $q$. As can be seen clearly, the average degree is increasingly higher under 2-Phase and it outperforms the random key pre-distribution as $q$ increases.

Figures 2–4 describe several sensornet security metrics. Figure 2(a) illustrates a measure of communication security (i.e invulnerability) by describing the average number of exclusive keys per pair of nodes in a cluster. This number is higher for nodes under 2-Phase than using the random scheme. Figure 2(b) measures the probability that a pair of nodes possesses at least one exclusive key under the 2-Phase key pre-distribution scheme. This probability rises sharply as $k$, the number of keys possessed by each node increases.

Figures 3 and 4 measure the vulnerability of communication links in a cluster under single as well as multiple node capture scenarios. As can be seen, the average number of links exposed to the adversary is lower under the 2-Phase scheme. The simulation results verify the analytical observations in Propositions 4 and 5 regarding link vulnerability. Lower link vulnerability for 2-Phase is explained by the fact that it is highly unlikely for captured nodes to have an LID adjacent to the LIDs of the communicating nodes.

## VIII. IMPLEMENTATION ISSUE: CREATING SORTED SHARED KEY LISTS

The security of a communication link strengthens with the exclusivity of the key(s) used for encryption on this link. For mutual communication each pair of nodes must therefore use keys shared among least number of nodes. During the shared key discovery phase, each node discovers its logical neighbors i.e., the neighbors with whom it shares at least one key. We propose the following metric to evaluate each shared key from this point of view.

Let $k$ be a key shared between any two nodes $i$ and $j$ and let $S_{ij}(k)$ denote the set of nodes in the neighborhood of $i$ and $j$ which share key $k$. Therefore, the eligibility of this key $k$ with respect to the pair of nodes $i$ and $j$ is defined as:

$$E_{ij}(k) = \begin{cases} 1 & \text{if } S_{ij}(k) = \phi \\ \frac{1}{|S_{ij}(k)|} & \text{otherwise} \end{cases} \qquad (14)$$

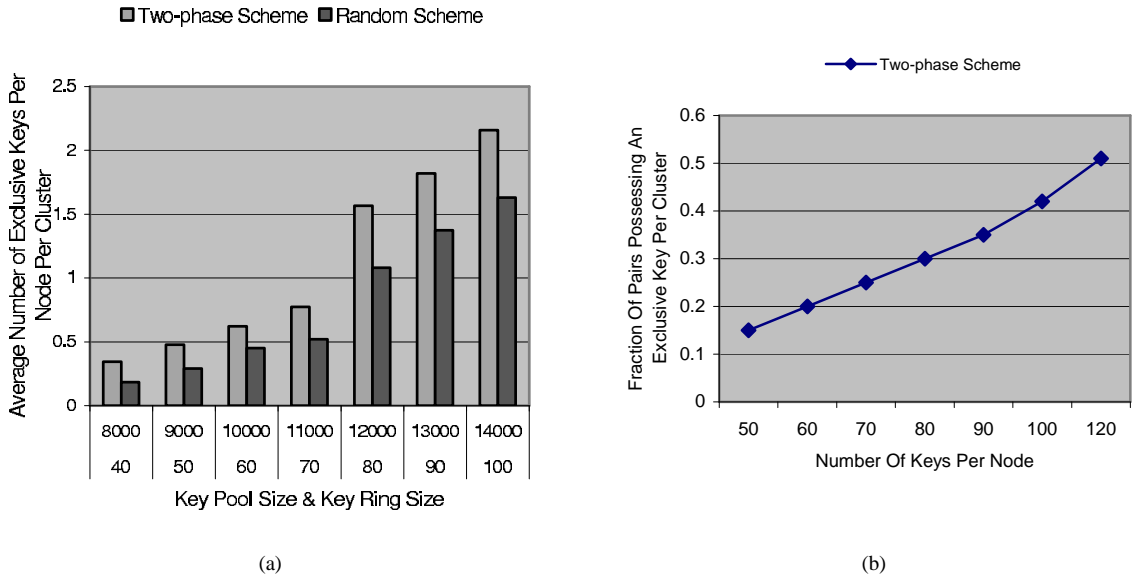

(a)                                    (b)

Fig. 2

(A) Av. # EXCLUSIVE KEYS PER NODE PAIR. (B) PROB. A NODE PAIR HAS AN EXCLUSIVE KEY.

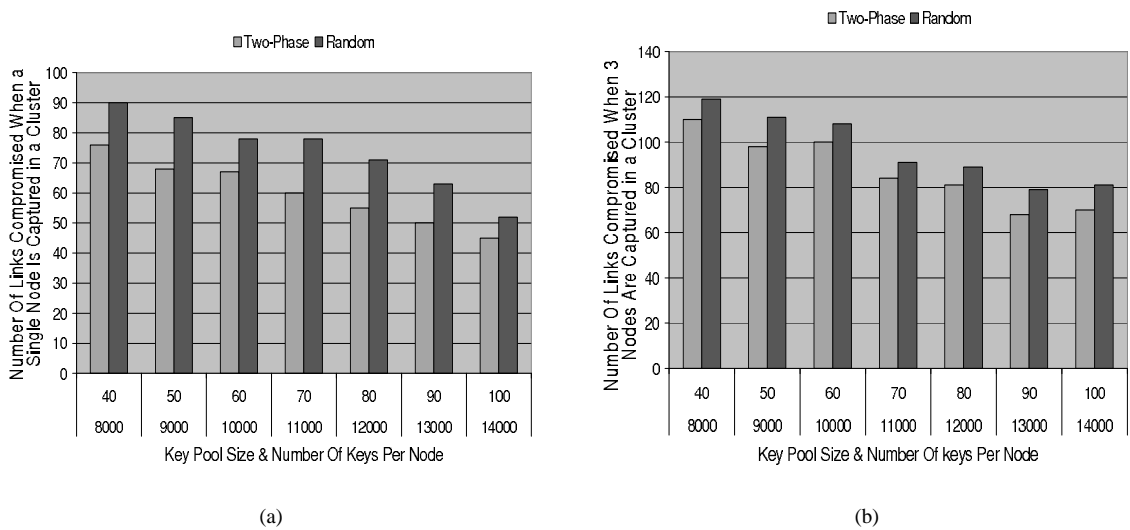

(a)                                    (b)

Fig. 3

AVERAGE # LINKS COMPROMISED IN A CLUSTER WHEN (A) ONE (B) THREE NODES ARE CAPTURED.

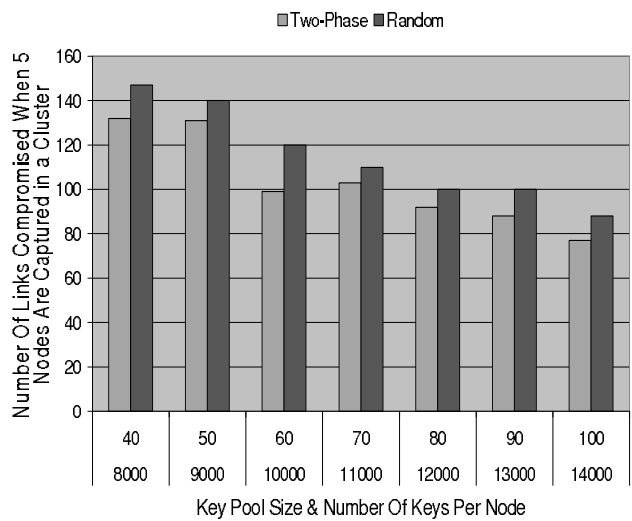

Fig. 4

AVERAGE # LINKS COMPROMISED IN A CLUSTER WHEN FIVE NODES ARE CAPTURED

The higher the value of $E_{ij}(k)$, the better is the key $k$ for communication between $i$ and $j$. During the shared key discovery phase, each node broadcasts the list of identifiers of the keys it possesses. Each node then create a separate list of shared keys for each of its neighbors sorted according to their eligibility values. The most eligible key should be used for communication until it is revoked.

## IX. CONCLUSION

Efficient pre-distribution of keys to sensor nodes is a very important issue for secure communication in sensor networks. Connectivity and resiliency to enemy attacks must be traded off very carefully. In this paper, we present an analytical framework with several quantitative metrics for evaluating key predistribution schemes and determining their security-performance tradeoff. We also present a 2-Phased key predistribution scheme based on a combination of inheritance and randomness which is proved to have better tradeoffs.